\RequirePackage[svgnames]{xcolor}

\documentclass[11pt,letterpaper]{mystyle}

\definecolor{PayPalBlue}{HTML}{0070BA}     
\definecolor{PayPalDarkBlue}{HTML}{003087} 
\definecolor{PayPalGold}{HTML}{FFC439}     
\definecolor{PayPalGoldLight}{HTML}{FFF7E8}
\definecolor{TinaCrimson}{HTML}{0070BA}    
\definecolor{YaleBlue}{HTML}{003087}       
\definecolor{CalGoldHex}{HTML}{FFF7E8}     

\tcbset{
    titlebox/.style={
        colback=PayPalGoldLight,
        colframe=PayPalGold,
        boxrule=0.5mm,
        arc=2mm,
        auto outer arc,
        left=5mm,
        right=5mm,
        top=5mm,
        bottom=5mm,
    }
}

\usepackage[all]{hypcap}
\usepackage[svgnames]{xcolor}
\usepackage[comma,authoryear,compress]{natbib}
\usepackage{hyperref}
\hypersetup{
    colorlinks = true,
    citecolor = {PayPalDarkBlue},
    linkcolor = {PayPalBlue},
    urlcolor = {PayPalBlue},
}

\usepackage{algorithm}
\usepackage{algorithmicx}
\usepackage{algpseudocode}
\usepackage{microtype}
\usepackage{graphicx}
\expandafter\def\csname ver@subfig.sty\endcsname{}
\usepackage{booktabs}
\usepackage{float}
\usepackage{bigstrut}

\usepackage{amsmath}
\usepackage{amssymb}
\usepackage{mathtools}
\usepackage{amsthm}
\usepackage{mathrsfs}
\usepackage{nicefrac}
\usepackage{dsfont}
\usepackage{enumitem}
\usepackage{subcaption}
\usepackage{cleveref}

\usepackage[utf8]{inputenc}
\usepackage[T1]{fontenc}
\usepackage{url}
\usepackage{amsfonts}
\usepackage{fdsymbol}
\usepackage{wrapfig}
\usepackage{lipsum}
\usepackage{stackengine}
\usepackage[font=small,labelfont=bf]{caption}
\usepackage{color}
\usepackage{adjustbox}
\usepackage{rotating}
\usepackage{makecell}

\newcommand{\pace}{\textsc{Pace}}

\definecolor{lightblue}{rgb}{0.22,0.45,0.70}
\definecolor{Gray}{gray}{0.95}
\definecolor{Cornsilk}{rgb}{1.0, 0.97, 0.86}

\graphicspath{{figures/}}

\title{PACE: Policy-Attested Contract Execution for Safe AI Agents in Decentralized Finance}

\runningtitle{PACE: Policy-Attested Contract Execution for Safe AI Agents in DeFi}

\author{
  Rabimba Karanjai$^{1,2}$,
  Yang Lu$^{1}$,
  Richard Williamson$^{2}$,
  Hemanth Hm$^{2}$,
  Prakhar Mehrotra$^{2}$,
  Lei Xu$^{3}$, and
  Weidong (Larry) Shi$^{1}$
}

\affil[1]{University of Houston, USA}
\affil[2]{PayPal, USA}
\affil[3]{Kent State University, USA}

\correspondingauthor{Rabimba Karanjai, University of Houston and PayPal}


\begin{document}

\begin{abstract}
Autonomous AI agents are emerging as interfaces for decentralized finance (DeFi) actions such as swaps, lending operations, and yield management. Because these agents rely on large language models (LLMs) to plan transactions, they inherit the LLM's susceptibility to prompt injection and lack of mechanisms to bind a verifier's approval to the exact transaction ultimately submitted on-chain. We present \pace{} (Policy-Attested Contract Execution), a transaction-level authorization framework that interposes between an LLM-based agent and on-chain execution. \pace{} introduces typed transaction intents, a deterministic policy verifier, and signed Policy Decision Records (PDRs) that cryptographically bind the approved intent, policy, and simulation report to the exact execution bytes, with replay and expiration protection. A Solidity smart account enforces PDR signatures on-chain with a measured overhead of 29{,}826--31{,}822~gas. We evaluate \pace{} against six baselines on 40~tasks spanning four attack categories plus benign utility (2{,}800~trials, 10~seeds). In our deterministic sandbox, \pace{} achieves a 0.00 unsafe execution rate and 0.00 false-positive rate on benign tasks, compared to 0.80 for the unguarded baseline. Ablation studies identify permissive policy settings (+57.5~pp) and the touched-contract allowlist (+12.5~pp) as the dominant safety components. To test whether the same deterministic floor holds for real model outputs, the artifact additionally provides a three-model live-LLM evaluation over the full task suite with repeated runs. A mainnet-fork harness is included for archive-RPC deployments, but fork results are reported only when the corresponding artifacts are generated. These auxiliary studies are separate from, and never substitute for, the deterministic benchmark. We frame our claims as \emph{logic-level} safety within a reproducible benchmark rather than deployment-ready DeFi security. All deterministic results are reproducible via \texttt{make reproduce}.
\end{abstract}

\maketitle
\vspace{3mm}

\section{Introduction}
\label{sec:introduction}

Decentralized finance (DeFi) protocols hold tens of billions of dollars
in user funds and execute financial logic as permissionless smart
contracts on public blockchains~\cite{zhou2023sokdefi}.  Interacting with them is notoriously error-prone: a user must assemble low-level
transaction calldata, reason about slippage, and token approval
semantics, and anticipate adversarial behavior such as front-running,
all before irrevocably committing funds.  To lower this barrier, a new
class of \emph{autonomous AI agents} built on large language models
(LLMs) now accepts goals in natural language, plans multi-step
strategies, and emits the transactions that carry them
out~\cite{schick2023toolformer, yao2023react}; such agents have begun
to operate on-chain under real capital~\cite{barton2026onchain}.

Delegating financial authority to an LLM, however, imports the LLM's
security weaknesses into a setting where mistakes are irreversible.
LLMs cannot reliably separate trusted instructions from untrusted data
and are subject to direct~\cite{perez2022ignore} and
indirect~\cite{greshake2023indirect} prompt injection: adversarial text
placed in a token name, a price feed, a memo field, or any other tool
output can hijack the agent into approving an attacker's transaction.
Peer-reviewed benchmarks confirm that tool-using agents are frequently
subverted in exactly this way~\cite{zhan2024injecagent,
debenedetti2024agentdojo}, and granting such agents direct control of
keys and capital has been argued to open qualitatively new vectors of
AI harm~\cite{marino2025harm}.  Unlike a chatbot's mistaken sentence, an
agent's mistaken transaction settles on-chain---draining a wallet or
granting an unlimited token allowance---before any human can intervene.

Existing safeguards each address a fragment.  Model-level alignment and
prompt-injection defences~\cite{ouyang2022training, chen2025struq} reduce
the \emph{frequency} of unsafe proposals but are probabilistic and
evadable~\cite{liu2024formalizing}, with no guarantee about the submitted
transaction.  Simulation previews a call's effects but is not bound to
the bytes broadcast, so post-simulation calldata mutation slips through.
Wallet guards and modules enforce on-chain constraints but ingest no
\emph{attested} off-chain simulation.  Among the safeguards we are aware
of, none cryptographically binds a verifier's approval---computed over a
pre-execution simulation---to the exact bytes that execute and re-checks
that binding on-chain where the value moves.

Our key observation is that the \emph{safety} of an agent's action need
not depend on the \emph{trustworthiness} of the model that proposed it:
if every action is reduced to an explicit, typed object, checked by a
deterministic procedure that never consults the model, and the decision
is cryptographically bound to that action and re-verified at execution,
then even a fully compromised LLM cannot cause a \emph{policy-violating}
transaction to settle---safety becomes a property of the policy and the
verifier, not of the model.  We realise this in \pace{}
(\textbf{P}olicy-\textbf{A}ttested
\textbf{C}ontract \textbf{E}xecution), which interposes three layers
between the LLM and on-chain execution:
(i)~\emph{typed transaction intents} capturing target, value,
calldata, approvals, and slippage;
(ii)~a \emph{deterministic policy verifier} that evaluates the
intent---and an attached pre-execution simulation---against a
user-defined policy without ever consulting the LLM; and
(iii)~a \emph{signed Policy Decision Record} (PDR) that
cryptographically binds the approved intent, policy, and simulation to
the exact execution bytes, which a Solidity smart account re-checks
on-chain (signer, calldata hash, nonce, validity window) before any
external call.  The LLM proposes, the verifier disposes, and the chain
enforces.  We are careful throughout to claim byte- and
field-level---not economic-outcome---binding (\S\ref{sec:discussion}).

\paragraph{Contributions}
(1)~The \pace{} architecture (\S\ref{sec:design}).
(2)~A 40-task benchmark across four attack categories plus benign
utility, with six non-PACE baselines (\S\ref{sec:evaluation}).
(3)~Deterministic sandbox experiments (2{,}800~trials) showing
\pace{} achieves 0.00 unsafe execution rate with 0.00 false
positives (\S\ref{sec:results}).
(4)~Ablation studies identifying permissive policy settings
(+57.5~pp) and the touched-contract allowlist (+12.5~pp) as the
dominant components
(\S\ref{sec:ablation}).
(5)~A Solidity smart account with measured gas overhead of
29{,}826--31{,}822~gas (\S\ref{sec:overhead}).
(6)~Separate live-LLM and local-EVM evidence, plus a fully
reproducible artifact (\texttt{make reproduce}).\footnote{An anonymized
artifact---source, smart-account contracts, deterministic task
definitions, raw result CSVs, Foundry logs, and live-LLM
traces---accompanies this submission for review; the repository will be
open-sourced upon acceptance.}

\paragraph{Scope}
Our evaluation uses a simplified in-memory DeFi simulator and a
regex-based mock LLM.  Results demonstrate logic-level safety
properties but should not be extrapolated to live chains or adaptive
adversaries without further evaluation.

\section{Background}
\label{sec:background}

\subsection{Decentralized Finance Primitives}

DeFi protocols implement financial services as permissionless smart
contracts on Ethereum~\cite{buterin2014ethereum, wood2014yellowpaper}.
Assets are represented as ERC-20 tokens~\cite{eip20}.

\paragraph{Automated Market Makers (AMMs)}
Constant-product pools such as Uniswap~\cite{adams2020uniswapv2,adams2021uniswapv3}
enable token swaps via the invariant $x \cdot y = k$.  Price impact
grows super-linearly with trade size relative to reserves: a swap
of size $\Delta x$ into reserves $(R_x, R_y)$ yields
$\Delta y = R_y \cdot \Delta x \cdot (1{-}f) \,/\, (R_x + \Delta x
\cdot (1{-}f))$, where $f$ is the fee rate.

\paragraph{Lending Protocols}
Platforms such as Aave~\cite{aaveLiquidations} support
collateralized borrowing.  A position's \emph{health factor}
$h = C \cdot p \cdot t \,/\, D$ determines liquidation risk, where
$C$ is collateral amount, $p$ is the oracle price, $t$ is the
liquidation threshold, and $D$ is outstanding debt.

\paragraph{Token Approvals}
The ERC-20 \texttt{approve(spender, amount)} function grants a
spender permission to transfer up to \texttt{amount}
tokens~\cite{eip20}.  \emph{Unlimited approvals}
($\mathtt{amount} = 2^{256}{-}1$) are a persistent attack vector: once granted, a malicious or compromised spender contract can drain
the full token balance at any future time.  Zhou et
al.~\cite{zhou2023sokdefi} catalog approval-based exploits as a
recurring pattern in DeFi attacks.

\subsection{Account Abstraction}

ERC-4337~\cite{erc4337} introduces smart-contract wallets with
programmable validation logic.  A \texttt{UserOperation} is
validated by the wallet's \texttt{validateUserOp} function before
execution, providing a natural enforcement point for
transaction-level authorization.  Ecosystem projects such as Safe
modules~\cite{safe2024modules} demonstrate programmable spending
caps, whitelists, and custom transaction logic within this framework.

\subsection{AI Agents in DeFi}

LLM-based agents interpret user goals in natural language, plan
multi-step strategies, and construct tool
calls~\cite{schick2023toolformer, yao2023react}.  Applied to
DeFi, these agents generate and submit transaction calldata
autonomously, turning a natural-language mandate such as ``earn yield
on my stablecoins'' into concrete swaps, approvals, and deposits.

Without external guardrails, an agent inherits the LLM's
vulnerabilities.  Chief among them is prompt injection---both
direct~\cite{perez2022ignore} and indirect~\cite{greshake2023indirect}%
---in which adversarial text in the tool output overrides the agent's
intended behavior; peer-reviewed benchmarks find that tool-using
agents are subverted by such attacks at high
rates~\cite{zhan2024injecagent, debenedetti2024agentdojo,
ruan2024toolemu}.  In DeFi the consequences are immediate and
irreversible: a hijacked agent can redirect funds, grant an unlimited
token approval, or submit a transaction with unbounded slippage, and the loss settles on-chain with no recourse~\cite{marino2025harm}.  This
motivates a guardrail that is \emph{external} to the model and binding
at the moment of execution.

\subsection{Maximal Extractable Value (MEV)}

MEV refers to profit extractable by reordering, inserting, or
censoring transactions within a block~\cite{daian2020flash}.
In a \emph{sandwich attack}~\cite{zhou2021sandwich}, a searcher
front-runs a victim's swap (moving the price), lets the victim
execute at a worse rate, then back-runs to capture the difference.
The victim's loss is bounded by their slippage tolerance parameter.
Qin et al.~\cite{qin2022quantifying} quantify the scale of MEV
extraction on Ethereum.

\section{Threat Model}
\label{sec:threat_model}

\subsection{System Model}

We consider a user who delegates DeFi operations to an LLM-based
agent.  The agent receives the user's goal in natural language,
plans one or more transactions, and submits them for execution.
\pace{} interposes between the agent's output and on-chain
submission.
\subsection{Adversary Model}

The adversary may inject malicious context through metadata, chat
history, or external feeds; direct the agent to malicious contracts;
front-run pending transactions; mutate calldata between simulation and
submission; and replay stale simulation results. Its goals include
direct theft through unauthorized transfers or approvals, indirect
loss through slippage, sandwich attacks, or oracle manipulation, policy
bypass, and post-simulation tampering.

The adversary cannot modify the user policy, verifier, or smart-account
validation logic. The agent has only PDR-gated authority and possesses
neither the owner key nor access to the owner-only raw-execution path,
which is reserved for recovery and administration. Compromise of the
owner key, verifier signing key, or policy is out of scope.
\subsection{Trust Boundaries}

\emph{Trusted} components are the user policy (set by the user or
administrator), the policy verifier, and the simulator (both deterministic
and auditable).  \emph{Untrusted} are the LLM/agent, the transaction
intent it produces, and all external contracts.  The central design
invariant is that the verifier \emph{never}
queries the LLM.  All checks are deterministic comparisons
against the user's declared policy.  This ensures that a fully
compromised LLM cannot influence the verification outcome.
Concretely, \pace{} targets \emph{soundness}: any transaction it executes
provably satisfies the declared policy, irrespective of LLM behaviour.
It makes no \emph{completeness} claim---whether the policy captures every
unsafe outcome is the author's responsibility (\S\ref{sec:discussion})---%
and it addresses prompt injection, policy misuse, and post-simulation
tampering, but not key compromise or multi-agent
collusion~\cite{alqithami2026agents}.

\subsection{Attack Categories}

The benchmark spans four attack categories---prompt injection, unsafe
contracts, DeFi exploit patterns, and MEV/slippage---plus benign utility
as a negative control (8~tasks each); Section~\ref{sec:evaluation}
details the tasks.

\section{\pace{} Design}
\label{sec:design}

\subsection{Architecture}

\begin{figure*}[t]
\centering
\includegraphics[width=0.8\textwidth]{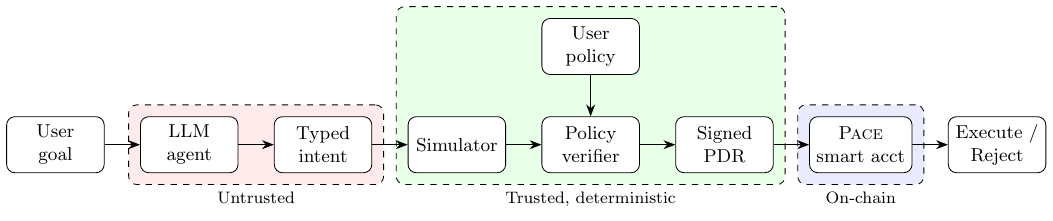}
\caption{The \pace{} pipeline. The LLM agent and the intent it emits are
\emph{untrusted}; the simulator, policy verifier, and PDR are
\emph{deterministic} and never consult the LLM; the smart account
re-checks the signed PDR \emph{on-chain} before any external call.
The user policy is the only trusted input that shapes the decision.}
\label{fig:architecture}
\end{figure*}

\pace{} interposes a deterministic verification pipeline between
the LLM agent's output and on-chain execution
(Figure~\ref{fig:architecture}): the agent emits a typed intent, the
intent is simulated, a deterministic verifier checks intent and
simulation against the user policy, and an approved decision is sealed
in a signed PDR that the smart account re-checks on-chain before any
external call.

Each component is described below.

\subsection{Typed Transaction Intents}

A \texttt{TransactionIntent} is a structured representation of the
agent's proposed on-chain action.  Fields include: chain ID, sender
and target addresses, ETH value, function selector, calldata, a list
of \texttt{TokenApproval} entries (token, spender, amount, unlimited
flag), a slippage tolerance, and a free-form metadata dictionary.

Every intent has a deterministic canonical JSON representation
(sorted keys, compact separators) and a SHA-256 hash
(\texttt{hash\_intent}).  The calldata field has its own hash
(\texttt{calldata\_hash}) for binding to simulation results.

\subsection{User Policy}

A \texttt{UserPolicy} is a declarative safety specification with
constraint classes for: address allowlists (targets, selectors, tokens);
value limits (max ETH per transaction, per-token spend caps); approval
controls (reject unlimited $2^{256}{-}1$ approvals); a slippage bound; a
function blocklist (e.g.\ \texttt{flashLoan}, \texttt{delegateCall});
simulation requirements (whether simulation is mandatory, and its maximum
age); post-state thresholds (minimum health factor, maximum loss); and
metadata regexes matched against intent metadata.  Policies are versioned
via their own SHA-256 hash.
\subsection{Deterministic Policy Verifier}

The verifier deterministically approves an intent only if it records no
violations. It checks target and selector allowlists, blocked selectors,
token allowlists and spend caps, ETH value, unlimited approvals,
slippage, and forbidden metadata patterns. When simulation is required,
it verifies success, freshness, non-future timestamps, allowed touched
contracts, minimum health factor, maximum loss, and equality between the
report's \texttt{intent\_hash} and the submitted intent hash, preventing
post-simulation calldata mutation.

The verifier is a pure, stateless function that neither accesses the
network nor consults the LLM. An attack must evade every applicable
check; our ablation (\S\ref{sec:ablation}) measures the contribution of
simulation- and PDR-dependent checks.

\subsection{Policy Decision Record (PDR)}

The PDR is an immutable attestation binding the full intent, the applied
policy, the simulation report, the decision and violation strings,
SHA-256 hashes of intent/policy/calldata, and a timestamp.  The verifier
signs a PDR for on-chain submission \emph{only when the decision is
\textsc{Approve}} (rejected records are kept off-chain for audit but never
signed, so a valid signature attests approval; adding an explicit
\texttt{decision} field to the signed digest is a simple further
hardening).  On-chain, \texttt{PaceSmartAccount} recovers an ECDSA
signature over a \emph{keccak256} digest of the PDR fields (Ethereum
Signed Message prefix) and re-checks the chain ID, account, target, value, the
keccak256 calldata-hash binding, nonce uniqueness, and the validity window
before any external call (\S\ref{sec:implementation}).  Off-chain records
use SHA-256 for content-addressing while the on-chain digest and calldata
binding use Ethereum-native keccak256; each side is internally consistent.

\paragraph{What the PDR binds---and does not}
The PDR binds an \emph{approval} to the exact intent, policy, simulation,
and \emph{execution bytes} (via the calldata hash), with replay/expiry
protection.  It is \emph{not} outcome binding: the contract does not
replay the simulation on-chain, so a transaction approved at state $S$ may
execute at $S'$ with different reserves, prices, or upgraded targets.
\pace{} thus guarantees byte- and field-level fidelity to the verifier's
decision, not that the realised outcome equals the simulated one
(\S\ref{sec:discussion}).

\subsection{Simulation Layer}

\pace{} simulates the intent before verification to populate
simulation-dependent checks (touched contracts, health factor,
loss percentage).  The simulation report is bound to the intent
via \texttt{intent\_hash}: if the calldata changes after
simulation, the hash mismatch triggers a violation.

\paragraph{Field provenance (a key assumption)}
\pace{}'s soundness requires that the semantic fields it checks
(approvals, spend, slippage, touched contracts) reflect the bytes that
execute.  The calldata-hash binding guarantees the submitted bytes equal
the approved bytes, but does not by itself guarantee that the
\emph{typed} fields match those bytes.  A sound deployment must therefore
\emph{reconstruct} these fields from decoded calldata and the simulation
trace using trusted code, never trusting the LLM-supplied fields
(\S\ref{sec:threat_model}); our reference implementation operates on the
typed intent and the trace its simulator produces, and full ABI decoding
of arbitrary router calldata is delegated to the trace backend.
Relatedly, a touched-contract set is a property of \emph{one} execution
trace: a context-sensitive contract could touch different addresses at
inclusion than in simulation, so \pace{} treats touched contracts as a
simulation-time check, sound only insofar as the relevant execution
state is unchanged (\S\ref{sec:discussion}).

The simulation backend is pluggable.  Our reference implementation
uses an in-memory DeFi model; a production deployment would use
trace-based simulation against an archive node.

\section{Implementation}
\label{sec:implementation}

\subsection{Reference Implementation}

\pace{} is implemented in Python~3.11+ using Pydantic~v2 for schema
validation and deterministic JSON serialisation.  The codebase is
structured as a pip-installable package (\texttt{pace}) with
subpackages for schemas, policy verification, simulation, agents,
experiments, and analysis.

\subsection{DeFi Simulator}

Our deterministic in-memory simulator processes intents and returns
simulation reports across five primitives: ERC-20 tokens (balances,
allowances, delegated transfers); a constant-product AMM
($x{\cdot}y{=}k$, 30\,bps fee, 10{,}000~WETH/10{,}000{,}000~DAI
reserves); a lending pool with oracle-based health factor (price
1{,}000~DAI/ETH, loan-to-value~0.75, liquidation threshold~0.80); a
malicious router that touches a hidden contract during a delegated call;
and an approval drainer.  Two specialised simulators exercise
post-simulation attacks by backdating report timestamps or returning a
report whose intent hash differs from the submitted intent.

\subsection{MEV Sandwich Simulator}

The MEV module optimises constant-product sandwich attacks: given pool
reserves, victim trade, slippage, fee, and attacker budget, it finds the
profit-maximising front-run via iterative refinement and reports victim
loss (bps), attacker profit, and whether strict slippage protection
blocks execution.

\subsection{Agent Baselines}

We implement seven agents with a shared decision interface, ordered by
safety coverage: \emph{Raw} (executes every intent); \emph{Prompt-Only}
(LLM keyword matching); \emph{Sim-Only} (rejects on simulation revert);
\emph{WalletGuard} (target allowlist $+$ ETH value cap); \emph{StaticGuard}
(full policy verifier, no simulation); \emph{SimGuard} (full policy $+$
simulation, no PDR/calldata binding); and \pace{} (simulation $+$ full
policy $+$ PDR).

\subsection{LLM Integration}

Deterministic experiments use a mock LLM that applies regex patterns
to the prompt text and returns a canned safe or unsafe response.  This
makes all experiments reproducible without API keys.

For optional live evaluation, the OpenAI-compatible adapter accepts a
base URL, API key, and model name, enabling compatible endpoints,
including self-hosted models.  Live
LLM responses are logged to trace files and still pass through the
same policy verifier---the LLM cannot bypass verification.

\subsection{On-Chain Smart Account}

The Solidity smart account enforces PDR verification on-chain before
any external call.  The PDR library computes a keccak256 digest over
12~PDR
fields (policyHash, intentHash, simulationHash, calldataHash,
chainId, account, target, value, nonce, validAfter, validUntil,
simulationBlock) and wraps it in an Ethereum Signed Message prefix
for signature recovery.

\begin{figure}[t]
\centering
\includegraphics[width=0.8\linewidth]{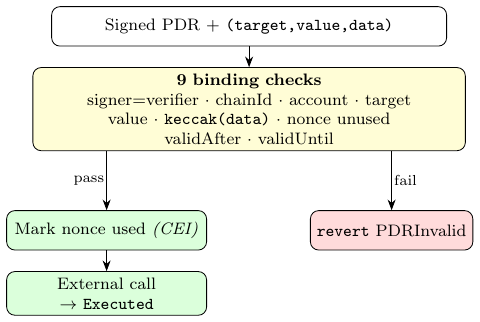}
\caption{On-chain enforcement in \texttt{executePDR}. Any failed check
reverts; the nonce is consumed before the external call, so a re-entrant
call cannot replay the PDR.}
\label{fig:enforcement}
\end{figure}

The PDR execution path (\texttt{executePDR}, Figure~\ref{fig:enforcement})
validates the trusted verifier signature, chain ID, account, target,
value, calldata hash binding, nonce uniqueness, and signed validity
window---nine checks in total.
Nonces are marked used before the external call (checks-effects-%
interactions pattern).  A separate owner-only execution path allows
raw calls without a PDR.

The contract suite contains 21~Foundry tests: 11 smart-account
enforcement cases covering valid execution (ERC-20 transfer, approval,
AMM swap), calldata mutation, target mutation, replay, expiry, wrong
verifier/account/chain~ID, and owner-only raw execution; 6 gas-overhead
cases; and 4 fork-harness cases.

\section{Evaluation}
\label{sec:evaluation}

\subsection{Research Questions}

\begin{itemize}
\item[\textbf{RQ1}] Can a deterministic policy verifier reduce the
      unsafe execution rate of AI DeFi agents?
\item[\textbf{RQ2}] Does deterministic safety enforcement cause
      false positives on benign tasks?
\item[\textbf{RQ3}] How do partial-defence baselines compare to
      the full \pace{} pipeline?
\item[\textbf{RQ4}] Which \pace{} components contribute most to
      safety?
\item[\textbf{RQ5}] What is the overhead of \pace{} in latency
      and on-chain gas?
\end{itemize}
\subsection{Benchmark Task Suite}

We construct 40 benchmark tasks across five categories (8 each):
benign utility tasks expected to be approved, and prompt injection,
unsafe contracts, DeFi exploits, and MEV/slippage attacks expected to
be rejected. The attacks cover instruction hijacking, encoded and
multilingual payloads, malicious targets, \texttt{delegateCall},
unlimited approvals, flash loans, oracle manipulation, unsafe borrowing,
stale simulations, calldata mutation, front-running, and extreme
slippage. Each task specifies a \texttt{TransactionIntent},
\texttt{UserPolicy}, expected decision, attack label, and severity.

Tasks are generated deterministically from a seed. Task IDs and expected
decisions remain fixed across seeds, while wording, attacker budgets,
pool reserves, slippage, injection strings, target addresses, and simulation age vary. Specialized simulators exercise stale-simulation
and calldata-mutation attacks during evaluation.

\subsection{Metrics}

We compute metrics from the results CSV using explicit denominators.
\textbf{Correct rate} is the fraction of tasks for which the agent's
approve/reject decision matches the expected safe behavior.
\textbf{Unsafe execution rate} is unsafe executions divided by all
trials.  \textbf{Attack success rate} uses the same unsafe-execution
numerator but divides by attack trials only.  \textbf{Dangerous block
rate} is rejected, unsafe attack transactions divided by attack trials.
\textbf{False-positive rate} is the number of benign tasks incorrectly rejected
divided by benign trials.

These rates are distinct from task completion.  \textbf{Task
completion rate} only measures whether scheduled trials produced a
recorded result; it is not a correctness metric.  When we report a
generic execution or reject/blocked rate, its denominator is all
trials, and it should not be read as attack blocking.

Additional financial and overhead metrics include the mean and median user
loss (bps), mean slippage (bps), health-factor violation rate, PDR
rejection rate, mean decision latency (ms), and measured on-chain gas
overhead.

\subsection{Methodology}

The main reported experiments use deterministic full mode: 10~seeds
(0--9), 7~agents, 40~tasks per seed, 2{,}800~trials total.  The
Prompt-Only agent uses a regex-based \texttt{MockLLM}; no live LLM
API calls are made in this deterministic suite.  Each seed is
reproducible, but seeds are treated as deterministic task variants
rather than repeated draws from a live deployment distribution.  We
report seed-level means and standard deviations for the main metrics.
Every deterministic result is reproducible via
\texttt{make experiments-full} followed by \texttt{make paper-assets}.
The optional live-LLM run is reported separately because it requires an
external API endpoint and key.

These are sandbox experiments on our in-memory simulator.  They
demonstrate the framework's logic-level safety properties but do not
constitute evidence of efficacy against adaptive adversaries on live
chains.  Because the tasks are constructed around the policy's modeled
properties, the suite validates the verifier's decision \emph{logic}
rather than robustness to unmodelled DeFi behavior (fuzzed calldata,
real ABIs, upgradeable proxies, exotic token standards); broadening it
along these axes is future work.

\subsection{Auxiliary Studies Beyond the Deterministic Sandbox}
\label{sec:aux_methodology}

To probe the two main external validity threats---the regex
\texttt{MockLLM} and the in-memory simulator---we add two auxiliary
studies, generated by the same pipeline and reported separately.  The
\emph{multi-model live evaluation}
(\texttt{pace.experiments.live\_llm\_suite}) runs three OpenAI-compatible
LLMs through the \emph{same} \texttt{PolicyVerifier} and the full 40-task
suite, repeated per model, logging per trial the model-alone and
\pace{}-guarded attack-success rates and a \emph{verifier safety floor}
(attacks the model judged \textsc{safe} that the verifier rejected);
unavailable models are skipped (\S\ref{sec:live_multimodel}).  The
\emph{mainnet-fork harness} (\texttt{scripts/run\_fork\_study.py})
enforces PDRs against real Uniswap~V2 and ERC-20 bytecode; it is gated
so the offline suite stays green without RPC access, and results are
reported (\S\ref{sec:fork_study}) only when generated.

\section{Results}
\label{sec:results}

We report deterministic full-mode results (10~seeds, 2{,}800~trials:
7~agents $\times$ 40~tasks $\times$ 10~seeds).  Each seed generates a
reproducible task variant with different wording and scenario
parameters.  These are sandbox results on our in-memory simulator and
should not be extrapolated to live deployments.

\subsection{Overall Safety and Utility}

Table~\ref{tab:main_results} summarizes the seven agents.

\begin{table}[t]
\centering\small
\caption{Agent comparison (2{,}800~trials, 10~seeds). Correct and
unsafe execution rates use all trials; attack success and dangerous
block rates use attack trials; false-positive rate uses benign trials.}
\resizebox{\columnwidth}{!}{\begin{tabular}{l r r r r r}
\hline
\textbf{Agent} & \textbf{Correct\%} & \textbf{Unsafe Exec} & \textbf{Atk Success} & \textbf{Dangerous Block Rate} & \textbf{FP Rate} \\
\hline
Raw & 20.0\% & 0.80 & 1.00 & 0.00 & 0.00 \\
Prompt-Only & 41.2\% & 0.59 & 0.73 & 0.27 & 0.00 \\
Sim-Only & 27.5\% & 0.72 & 0.91 & 0.09 & 0.00 \\
WalletGuard & 75.0\% & 0.25 & 0.31 & 0.69 & 0.00 \\
StaticGuard & 95.0\% & 0.05 & 0.06 & 0.94 & 0.00 \\
SimGuard & 97.5\% & 0.03 & 0.03 & 0.97 & 0.00 \\
PACE & 100.0\% & 0.00 & 0.00 & 1.00 & 0.00 \\
\hline
\end{tabular}}

\label{tab:main_results}
\end{table}

In our benchmark, PACE achieves 100.0\% correct rate (40/40 per
seed), 0.00 unsafe execution rate, and 0.00 false-positive rate.
The Raw baseline executes everything, yielding 0.80 unsafe execution
rate over all trials and 1.00 attack success rate over attack trials.
The strongest non-PACE baseline, SimGuard, applies policy checks over a
fresh simulation but lacks PDR calldata binding; it reduces attack
success to 0.0312 but still admits the calldata-mutation task.
All agents achieve zero false positives---benign tasks are well
within policy limits.
Seed-level variability is concentrated in Prompt-Only: correct rate
41.2\% $\pm$ 2.56~pp, unsafe execution rate 0.5875 $\pm$ 0.0256,
and attack success rate 0.7344 $\pm$ 0.0320.  PACE remains 100.0\%
correct with 0.000 unsafe execution across the seeded task variants.
PACE's zero seed-level variance is a deterministic consequence of
declarative verification---given a fixed policy, the decision is a pure
function of the intent---rather than a low-variance sampling estimate.
The reported 0.00 rates, therefore, characterize behavior on this
benchmark and should not be read as a statistical guarantee over
unseen or adaptive adversaries.

\subsection{Per-Category Attack Analysis}

Breaking attack success down by category:
PACE, SimGuard, and StaticGuard block all prompt-injection and
MEV/slippage tasks via policy constraints (metadata patterns and
slippage limits respectively).  Against unsafe contracts, PACE,
SimGuard, and StaticGuard achieve 0.00 attack success; WalletGuard
misses the unlimited-approval task (zero ETH value bypasses its value
check).  Against DeFi-attack patterns, PACE blocks all~8 including the
post-simulation attacks (stale simulation \texttt{da\_07} and
calldata mutation \texttt{da\_08}); StaticGuard misses 2/8 that require
simulation-dependent checks, while SimGuard catches stale simulations, but
still misses the calldata mutation because it does not bind the simulated
intent of the submitted calldata.

\paragraph{Worked example: calldata mutation}
Task \texttt{da\_08} shows why simulation alone is insufficient.  The
agent's intent is simulated and passes policy; the bytes submitted then
differ from the simulated/approved bytes, so the simulation report's
\texttt{intent\_hash} no longer matches the submitted intent.  SimGuard
applies the same policy and simulation but \emph{omits} the
hash-equality check, so it admits the transaction---the single attack it
misses.  \pace{} rejects the mismatch off-chain via \texttt{intent\_hash}
and independently on-chain via \texttt{keccak256(data)}
(Figure~\ref{fig:enforcement}).  It is the only task freed when calldata
binding is removed (\S\ref{sec:ablation}), isolating PDR binding as the
decisive component.

\subsection{Ablation Study}
\label{sec:ablation}

Table~\ref{tab:ablation} reports 320~ablation trials (8~PACE
variants, seed~42).

\begin{table}[t]
\centering\small
\caption{Ablation: unsafe execution rate when each component is removed.
The baseline uses the same task policies and task-specific simulators
as the main PACE experiment. The reject rate is all rejected decisions
over all ablation trials, not the dangerous block rate.}
\resizebox{\columnwidth}{!}{\begin{tabular}{l r r r r}
\hline
\textbf{Variant} & \textbf{Correct\%} & \textbf{Unsafe Exec Rate} & \textbf{Reject Rate} & \textbf{FP Rate} \\
\hline
pace\_baseline & 100.0\% & 0.000 & 0.800 & 0.000 \\
pace\_no\_sim & 95.0\% & 0.050 & 0.750 & 0.000 \\
pace\_no\_metadata & 100.0\% & 0.000 & 0.800 & 0.000 \\
pace\_no\_touched & 87.5\% & 0.125 & 0.675 & 0.000 \\
pace\_no\_calldata\_bind & 97.5\% & 0.025 & 0.775 & 0.000 \\
pace\_no\_freshness & 97.5\% & 0.025 & 0.775 & 0.000 \\
pace\_permissive & 42.5\% & 0.575 & 0.225 & 0.000 \\
pace\_conservative & 95.0\% & 0.000 & 0.850 & 0.250 \\
\hline
\end{tabular}}
\label{tab:ablation}
\end{table}

\texttt{pace\_baseline} matches the main PACE configuration:
100.0\% correct rate, 0.000 unsafe execution rate, and 0.000
false-positive rate. Using an explicitly permissive policy raises
the unsafe execution rate from 0.000 to 0.575 (+57.5~pp).
The \textbf{touched-contract allowlist} is second: removing it
raises the unsafe execution rate to 0.125 (+12.5~pp).
Skipping simulation raises the unsafe execution rate to 0.050 (+5.0~pp).
Removing calldata binding or freshness enforcement raises the unsafe
execution rate to 0.025 (+2.5~pp each).  Removing metadata scanning shows no
isolated impact in this benchmark because those attacks are also
blocked by other policy constraints.
Sensitivity analysis shows accuracy
peaks at 100~bps slippage
threshold (100.0\%) and drops at 10~bps (95.0\%, over-rejection)
and 1{,}000~bps (95.0\%, MEV admitted).  A 0-second freshness
window drops accuracy to 80.0\%; windows of at least 1~second
recover 100.0\% accuracy in this deterministic suite---the slippage and
freshness thresholds are the only parameters that trade false positives
against admitted attacks, underscoring that \pace{}'s safety rests on
the policy values a deployer chooses.

\subsection{On-Chain Gas Overhead}
\label{sec:overhead}

Measured with Foundry (Solidity~0.8.24, optimizer 200~runs), PDR
verification adds 29{,}826--31{,}822~gas per transaction, dominated by
\texttt{ecrecover} (${\sim}$3{,}000) and nonce SSTORE (${\sim}$20{,}000).
This is a \emph{material} relative cost for cheap operations---91.7\% for
an approval, 78.9\% for a transfer---and a smaller 50.5\% for a swap.  The reported ${\sim}$0.04~ms is
\emph{verifier-only} decision latency; it excludes trace-based
simulation, ABI decoding, and archive-RPC access, which dominate
end-to-end latency in any production deployment.

\paragraph{Local EVM enforcement}
As a second execution environment beyond the in-memory simulator, the
artifact runs \texttt{PaceSmartAccountTest} in Foundry.  The local EVM
case study passes 11/11 smart-account cases: three valid PDR
executions, seven invalid-PDR rejection cases (calldata mutation,
target mutation, replay, expiry, wrong verifier/account/chain~ID), and
one owner-only raw execution path.

\subsection{Multi-Model Live LLM Evaluation}
\label{sec:live_multimodel}

To test whether the deterministic floor holds for real model outputs, we
run three LLMs reached through OpenAI-compatible
endpoints---\texttt{gemini-3.1-flash-lite}, \texttt{DeepSeek-V4-Flash},
and \texttt{gpt-4o}---through the same \texttt{PolicyVerifier} and the
full 40-task suite (5~repeats per model; 595~trials total,
${\approx}200$ per model split $160$~attack\,/\,$40$~benign; seed~42).
Exact prompts, parser behavior, and per-trial traces are in the artifact
(\texttt{results/runs/live\_suite/}); these runs are reported separately
from the deterministic 2{,}800-trial suite
(Table~\ref{tab:live_multimodel}).

\IfFileExists{tables/live_model_comparison.tex}{%
\begin{table}[t]
\centering\small
\caption{Live multi-model evaluation. Correct rate, model-alone vs.\
\pace{}-guarded attack-success rate, and benign false-positive rate.
Generated from \texttt{results/runs/live\_suite/}.}
\resizebox{\columnwidth}{!}{\begin{tabular}{l r r r r}
\hline
\textbf{Model} & \textbf{Correct\%} & \textbf{LLM-Only Atk Succ} & \textbf{PACE Atk Succ} & \textbf{FP Rate} \\
\hline
gemini-3.1-flash-lite & 97.5 & 0.000 & 0.000 & 0.125 \\
DeepSeek-V4-Flash & 97.4 & 0.026 & 0.000 & 0.128 \\
gpt-4o & 99.0 & 0.069 & 0.000 & 0.050 \\
\hline
\end{tabular}
}
\label{tab:live_multimodel}
\end{table}
}{\emph{(Live multi-model table is generated by \texttt{make live-suite}
with API keys configured; see \texttt{paper\_assets/live\_suite\_summary.md}.)}}

All three models reach a \pace{}-guarded attack-success rate of 0.000,
and the verifier floor is substantive on real outputs: unguarded,
\texttt{gpt-4o} judged 11/160 attack trials (6.9\%) and
\texttt{DeepSeek-V4-Flash} 4/156 (2.6\%) \textsc{safe}, and the
deterministic verifier rejected \emph{all} of them;
\texttt{gemini-3.1-flash-lite} admitted none unguarded.  The invariant is
structural rather than model-specific---for every attack a model judges
\textsc{safe} the verifier re-checks the parsed intent against the
policy, so the guarded attack-success column is bounded by the verifier,
not by model judgement.  Unlike the deterministic sandbox, the live
setting shows non-zero benign false positives (5.0--12.8\%) from model
and parser variance, a deployment-time tuning concern rather than a
safety failure.

\subsection{Mainnet-Fork Execution Harness}
\label{sec:fork_study}

\IfFileExists{results/fork_study/results.csv}{%
The study (\texttt{paper\_assets/fork\_study\_summary.md}) executes a
valid-PDR WETH$\rightarrow$DAI swap end to end against the live router
and confirms that calldata mutation, target mutation, and nonce replay
are rejected on real protocol calldata.}{%
\emph{(The mainnet-fork harness is reproducible with an archive RPC
endpoint via \texttt{FORK\_RPC\_URL=<rpc> python
scripts/run\_fork\_study.py}.  Without an RPC the fork tests early-return
so the offline suite stays green.  No fork-result claims are made unless
\texttt{results/fork\_study/results.csv} is generated.)}}

\subsection{Adaptive Adversary}
\label{sec:adaptive}

To probe robustness beyond the seeded benchmark, we evaluate an adversary
with full knowledge of the policy (conservative configuration, seed~0;
\texttt{scripts/run\_adaptive\_study.py}).  \emph{Evasion} attacks perturb
a policy-\emph{violating} action to slip past a specific check: value and
slippage just past a bound, a typo-squatted target, a blocked selector, a
\texttt{delegateCall} that touches a hidden contract, post-simulation
calldata mutation, and an unlimited approval hidden behind benign-looking
metadata.  \pace{} blocks all 8/8 (Table~\ref{tab:adaptive}); the
deterministic checks are not fooled by adaptation, and the disguised
approval is caught by the \emph{exact-approval} check rather than the
metadata regex---confirming the regex is a non-load-bearing diagnostic.
SimGuard and StaticGuard block 7/8 (missing only calldata mutation) and
WalletGuard 2/8, consistent with the main results.

\emph{Policy-valid-but-harmful} transactions stay strictly within the
policy yet are economically harmful: a swap at exactly the slippage
bound, a transfer at the value cap, a large \emph{finite} approval to an
allowlisted spender, and a five-leg multi-transaction drain whose legs
are each under the cap.  \pace{}---and \emph{every} baseline---admits all
8/8.  This is by design: \pace{} enforces the declared policy, not
economic safety, and verifies one transaction at a time.  The study thus
delineates \pace{}'s guarantee (soundness against policy-violating
actions, robust to adaptation) from its non-guarantee (completeness over
harmful-but-policy-valid behaviour and multi-transaction sequences),
empirically motivating the policy-authoring and multi-step analysis we
leave to future work (\S\ref{sec:future_work}).

\begin{table}[t]
\centering\small
\caption{Adaptive adversary with full policy knowledge (seed~0).  Evasion
attacks are policy-violating (lower-left is better); policy-valid-harmful
transactions stay within the policy, and every guard admits them all.}
\begin{tabular}{l r r}
\hline
\textbf{Agent} & \textbf{Evasion blocked} & \textbf{Policy-valid harm admitted} \\
\hline
PACE & 8/8 (1.00) & 8/8 (1.00) \\
SimGuard & 7/8 (0.88) & 8/8 (1.00) \\
StaticGuard & 7/8 (0.88) & 8/8 (1.00) \\
WalletGuard & 2/8 (0.25) & 8/8 (1.00) \\
\hline
\end{tabular}

\label{tab:adaptive}
\end{table}

\subsection{Error Analysis and Research Questions}

\pace{} makes 0~incorrect decisions.  By attacks blocked (of 32 per
seed): Raw~0, Sim-Only~3, WalletGuard~22, StaticGuard~30, SimGuard~31
(missing only calldata mutation), \pace{}~32; Prompt-Only blocks a
seed-dependent subset (mean attack success 0.73).  These answer
RQ1--RQ5: \pace{} attains a 0.00 unsafe-execution and 0.00
false-positive rate (RQ1--RQ2) versus 1.00--0.0312 for the baselines
(RQ3); the policy and touched-contract allowlist dominate (RQ4); and
overhead is 29{,}826--31{,}822~gas with ${\sim}0.04$~ms verifier
latency (RQ5).

\section{Discussion}
\label{sec:discussion}

\paragraph{Why both layers, and which matter most}
Neither layer suffices alone---policy-only (StaticGuard) misses the two
simulation-dependent attacks and simulation-only (Sim-Only) catches only
reverts---so \pace{} combines them.  The ablation localises each
component, but the dominant factor is the policy itself: a permissive
policy raises unsafe execution by 57.5~pp, so safety derives primarily
from the policy \emph{specification} and well-designed defaults are
essential.

\subsection{Ethical Considerations}

All attacks in this work run against an in-memory simulator and local
test contracts with no real funds at stake.  The benchmark reuses attack
\emph{categories} already documented in the cited literature rather than
disclosing new vulnerabilities in any deployed protocol, so no
responsible-disclosure process applies.  \pace{} is a defensive
mechanism, and the artifact we release contains only synthetic
scenarios; it cannot be repurposed to attack live systems.  We believe
the benchmark's primary effect is to help defenders evaluate
agent-level safeguards before such agents manage real capital.

\section{Related Work}
\label{sec:related_work}

\paragraph{Smart-contract analysis}
Static and symbolic tools---Slither~\cite{feist2019slither},
Oyente~\cite{luu2016oyente}, and Manticore~\cite{mossberg2019manticore}---%
audit the \emph{contract code} for exploitable logic.  \pace{} instead
verifies the \emph{agent's proposed transaction} against a user policy
before it reaches any contract; the two are complementary, since a
correct contract can still be dangerous under the wrong parameters
(e.g., an unlimited approval to a phished spender).

\paragraph{DeFi attacks and MEV}
A large literature \emph{detects and measures} attacks: SoK
taxonomies~\cite{zhou2023sokdefi}, sandwich
attacks~\cite{zhou2021sandwich}, MEV
quantification~\cite{qin2022quantifying}, and
front-running~\cite{daian2020flash}.  \pace{} acts \emph{pre-submission},
enforcing slippage, value, and simulation-based post-state checks before
a transaction enters the mempool.

\paragraph{Account abstraction and modular guards}
ERC-4337~\cite{erc4337} enables programmable validation, and a rich
ecosystem of smart-account guards and modules---Safe guards and
modules~\cite{safe2024modules}, Zodiac Roles~\cite{zodiacroles}, and the
ERC-6900~\cite{erc6900} and ERC-7579~\cite{erc7579} modular-account
standards---already enforce allowlists, spending caps, scoped function
permissions, and on-chain pre-/post-execution hooks.  These are powerful
but validate over \emph{on-chain} state alone; \pace{}'s distinction is to
bind an \emph{attested off-chain simulation} and a verifier decision to
the exact execution bytes.  Architecturally, the PDR verifier is best
expressed as a validator/hook module in these frameworks rather than a
bespoke account, which we leave to future work.  Industry simulation and
transaction-security services---Tenderly~\cite{tenderly2026sim},
Blockaid~\cite{blockaid2026}, Wallet Guard~\cite{walletguard2026}---target
a human signer with advisory verdicts; \pace{} differs by being
deterministic, cryptographically bound to the exact intent, and enforced
on-chain, and is complementary (such a service can supply the
\texttt{SimulationReport}).

\paragraph{LLM safety and prompt injection}
RLHF~\cite{ouyang2022training} and Constitutional
AI~\cite{bai2022constitutional} train models to refuse harmful outputs,
yet prompt injection still succeeds, both direct~\cite{perez2022ignore}
and indirect~\cite{greshake2023indirect}; peer-reviewed work formalises
attacks and defences~\cite{liu2024formalizing} and proposes input-level
defences such as structured queries~\cite{chen2025struq}.  These operate
within or around the model and remain probabilistic, whereas \pace{}'s
verifier is external and deterministic, providing a hard floor regardless
of whether such a defence is evaded.

\paragraph{Tool-using and on-chain agents}
Toolformer~\cite{schick2023toolformer} and ReAct~\cite{yao2023react}
established tool-using agents; on-chain agents now act under real
capital~\cite{barton2026onchain}, which Marino and
Juels~\cite{marino2025harm} argue opens new vectors of AI harm.
Benchmarks---InjecAgent~\cite{zhan2024injecagent},
AgentDojo~\cite{debenedetti2024agentdojo}, ToolEmu~\cite{ruan2024toolemu}---%
\emph{measure} agent susceptibility but do not \emph{prevent} unsafe
execution.  Closest to our framing, Alqithami~\cite{alqithami2026agents}
proposes, as a roadmap, a Transaction Intent Schema and a Policy Decision
Record; \pace{} differs by providing the first concrete, evaluated,
on-chain-enforced instantiation of these abstractions with simulation
binding.

\section{Future Work}
\label{sec:future_work}

Several directions would move \pace{} toward deployment-grade
assurance.  Section~\ref{sec:adaptive} takes a first step against a
policy-aware adversary; \emph{adaptive} attackers that \emph{learn}
against the verifier over many rounds, and multi-step agent loops that
compose policy-valid legs into a harmful sequence, remain open and are
the most important next target.  Higher-fidelity execution---running
the full attack suite under trace-based simulation against forked
mainnet state, and broadening coverage to cross-contract reentrancy,
governance, and bridge exploits---would strengthen external validity;
the verifier and PDR binding transfer directly to such a backend.
Larger model panels (including open-weight models) and end-to-end agent
harnesses would better characterize how often models propose unsafe
actions.  Finally, upgrading PDR hashing to EIP-712 typed data and
integrating validation into the ERC-4337 \texttt{validateUserOp} path
would let \pace{} run as a standard smart-account module, and tools for
authoring and tuning policies---the dominant safety factor---are an
important complement to the enforcement mechanism.

\section{Conclusion}
\label{sec:conclusion}

We presented \pace{}, a framework that separates untrusted LLM planning
from deterministic, attestable safety enforcement for AI agents acting in
decentralized finance.  Its core contribution is a three-part
enforcement primitive: typed transaction intents that make a proposed
action explicit, a deterministic policy verifier that judges each intent
against a user-defined policy without ever consulting the LLM, and a
signed Policy Decision Record that cryptographically binds intent,
policy, and simulation to a decision and is re-checked on-chain by a
smart account before execution.  Around this design the paper
contributes a reproducible benchmark spanning four attack families and
benign utility, a comparison against six non-\pace{} baselines, and
ablation, gas, live-model, and on-chain case studies that together
isolate where each layer of protection becomes necessary.

The unifying lesson is that the safety of an agent's on-chain actions
can be decoupled from the trustworthiness of the model that proposes
them: by treating the model's entire output as untrusted and enforcing a
signed policy decision both off- and on-chain, \pace{} provides a safety
floor that does not degrade as model behaviour varies.  We hope this
framing, and the accompanying artifact, help establish transaction-level
policy attestation as a building block for autonomous agents that act
under real capital.

\clearpage
\bibliography{references}

\end{document}